\documentclass[aps,prl,preprint,groupedaddress,notitlepage,tightenlines]{revtex4-1}

\usepackage{amsmath}
\usepackage{graphicx}

\preprint{NuFact15 - Rio de Janeiro, Brazil - August 2015}

\begin{document}

\newcommand{\peii}{$\pi_{e2}$}
\newcommand{\peiig}{$\pi_{e2\gamma}$}
\newcommand{\peiiee}{$\pi_{e2ee}$}
\newcommand{\peiii}{$\pi_{e3}$}
\newcommand{\VmA}{$V$$-$$A$}


\title{Allowed rare pion and muon decays as tests of the Standard Model}
\thanks{\it Presented at NuFact15, 10-15 Aug 2015, Rio de Janeiro, 
Brazil [C15-08-10.2]}



\author{Dinko Po\v{c}ani\'c}
\email[]{pocanic@virginia.edu}
\thanks{Speaker}
\affiliation{Institute for Nuclear and Particle Physics, University of
  Virginia, Charlottesville, VA 22904-4714, U.S.A.}

\collaboration{PEN Collaboration}
\thanks{L.\,P.~Alonzi,
        V.\,A.~Baranov,
        W.~Bertl,
        M.~Bychkov,
        Yu.\,M.~Bystritsky,
        E.~Frle\v{z},
        C.\,J.~Glaser,
        V.\,A.~Ka\-lin\-nik\-ov,
        N.\,V.~Khomutov,
        A.\,S.~Korenchenko,
        S.\,M.~Korenchenko,
        M.~Korolija,
        T.~Koz\-low\-ski,
        N.\,P.~Kravchuk,
        N.\,A.~Kuchinsky,
        M.\,C.~Lehman,
        D.~Mekterovi\'c,
        E.~Munyangabe, 
        D.~Mzhavia,
        A.~Palladino,
        P.~Robmann,
        A.\,M.~Rozhdestvensky,
        I.~Supek,
        P.~Tru\"ol,
        Z.~Tsamalaidze,
        A.~van~der~Schaaf,
        B.~Vandevender,
        E.\,P.~Velicheva,
        M.\,G.~Vitz,
        V.\,P.~Volnykh.
          }
\noaffiliation

\date{20 December 2015}

\begin{abstract}
Simple dynamics, few available decay channels, and extremely well
controlled radiative and loop corrections, make pion and muon decays a
sensitive means for testing the underlying symmetries, the universality
of weak fermion couplings, as well as for study of pion structure and
chiral dynamics.  We review the current state of experimental study of
the allowed rare decays of charged pions: (a) electronic, $\pi^+ \to
e^+\nu_e$, or $\pi_{e2}$, (b) radiative, $\pi^+ \to e^+\nu_e\gamma$, or
$\pi_{e2\gamma}$, and (c) semileptonic, $\pi^+\to \pi^0 e^+ \nu$, or
$\pi_{e3}$, as well as muon radiative decay, $\mu^+\to e^+
\nu_{\text{e}}\bar{\nu}_{\mu}\gamma$.  Taken together, these data
present an internally consistent picture that also agrees well with
Standard Model (SM) predictions.  However, even following the great
strides of the recent decades, experimental accuracy is lagging far
behind that of the theoretical description for all above processes.  We
review the implications of the present state of knowledge and prospects
for further improvement in the near term.
\end{abstract}

\pacs{13.20.Cz,13.35.Bv}
\keywords{Pion decays, muon decays, lepton universality}

\maketitle

\section{Motivation}

Pion decay has provided an important testing ground for the weak
interaction and radiative corrections from the beginnings of modern
subatomic physics.  The unexpected suppression of the direct electronic
decay of the pion ($\pi\to e\nu$, or \peii) led to an early examination
of the nature of the weak interaction and to the prediction of a low
branching fraction of $\sim 1.3\times 10^{-4}$ \cite{Fey58} as a
consequence of the \VmA\ nature of the weak interaction, through
helicity suppression of the right-handed state of the electron.  In the
meantime, the extraordinary success of the Standard Model has opened
significant opportunities for precision tests of its underlying
symmetries, lepton and quark-lepton coupling universality, and a host of
related issues through precision measurements of pion decays.  We will
address the specific motivation and physics reach for each channel
separately below.  A recent in-depth review of the subject is given in
Ref.~\cite{Poc14}. 

Muon decay, a purely leptonic electroweak process, serves a special role
in the Standard Model because it calibrates the strength of the weak
coupling.  Its precise theoretical description, via the so-called Michel
parameters \cite{Mic50}, positions it uniquely to provide constraints on
possible contributions outside the \VmA\ standard electroweak model.
Below we discuss new results on the muon radiative decay $\mu^+ \to e^+
\nu_{e} \bar{\nu}_\mu\gamma$, the only process that gives access to the
decay parameter $\bar{\eta}$.

\section{\boldmath Pion electronic $\pi\to e\bar{\nu}$ decay (\peii)
      \label{sec:pi_e2}}

At the tree level, the ratio of the $\pi \to e\bar{\nu}_e$ to $\pi \to
\mu\bar{\nu}_\mu$ decay widths is given by \cite{Fey58,Bry82}
\begin{equation}
    R_{e/\mu,0}^\pi = \frac{\Gamma(\pi \to  e\bar{\nu}_e)}
          {\Gamma(\pi \to  \mu\bar{\nu}_\mu)}
       = \frac{m_e^2}{m_\mu^2}\cdot
        \frac{(m_\pi^2-m_e^2)^2}{(m_\pi^2-m_\mu^2)^2}
      \simeq 1.283 \times 10^{-4}\,,  \label{eq:pi_e2_tree}
\end{equation}
where the ratio of squared lepton masses for the two decays, comes from
the helicity suppression by the \VmA\ lepton-$W$ boson weak couplings.
If, instead, the decay could proceed directly through the pseudoscalar
current, the ratio $R_{e/\mu}^\pi$ would reduce to the second,
phase-space factor, or approximately 5.5.  More complete treatment of
the process includes $\delta R_{e/\mu}^\pi$, the radiative and loop
corrections, and the possibility of lepton universality violation, i.e.,
that $g_e$ and $g_\mu$, the electron and muon couplings to the $W$,
respectively, may not be equal:
\begin{equation}
    R_{e/\mu}^\pi =\frac{\Gamma(\pi \to  e\bar{\nu}(\gamma))}
          {\Gamma(\pi \to  \mu\bar{\nu}(\gamma))}
      = \frac{g_e^2}{g_\mu^2}\frac{m_e^2}{m_\mu^2}
        \frac{(m_\pi^2-m_e^2)^2}{(m_\pi^2-m_\mu^2)^2}
           \left(1+\delta R_{e/\mu}^\pi \right)\,,
    \label{eq:pi_e2_general}
\end{equation}
where the ``$(\gamma)$'' indicates that radiative decays are fully
included in the branching fractions.  Improvements of the theoretical
description of the $\pi_{e2}$ decay have culminated in a series of
calculations that have refined the SM prediction to a precision of 8
parts in $10^5$:
\begin{equation}   \label{eq:pi_e2_full_SM}
     \left(R_{e/\mu}^{\pi}\right)^{\rm SM} = 
       \frac{\Gamma(\pi \to e\bar{\nu}(\gamma))}
          {\Gamma(\pi \to  \mu\bar{\nu}(\gamma))}\bigg|_{\text{calc}} =
   \begin{cases}
    1.2352(5) \times 10^{-4} & \text{Ref.~\cite{Mar93},} \\
    1.2354(2) \times 10^{-4} & \text{Ref.~\cite{Fin96},} \\
    1.2352(1) \times 10^{-4} & \text{Ref.~\cite{Cir07}.} 
   \end{cases}
\end{equation}
A comparison with equation (\ref{eq:pi_e2_tree}) reveals that the
radiative and loop corrections amount to almost 4\% of
$R_{e/\mu}^{\pi}$.  However, the current experimental precision
\cite{PDG14,Agu15} 
\begin{equation} \label{eq:pi_e2_2015_avg}
  \left(R_{e/\mu}^{\pi}\right)^{\text{EXP}} = 
    1.2327\,(23) \times 10^{-4}\,,
\end{equation}
lags behind the theoretical one by more than an order of magnitude.

Because of the large helicity suppression of the \peii\ decay, its
branching ratio is highly susceptible to small non-\VmA\ contributions
from new physics, making this decay a particularly suitable subject of
study, as discussed in, e.g.,
Refs.~\cite{Sch81,Sha82,Loi04,Ram07,Cam05,Cam08}.  This prospect
provides the primary motivation for the ongoing PEN \cite{PENweb} and
PIENU \cite{PiENuWeb} experiments.  Of the possible ``new physics''
contributions in the Lagrangian, \peii\ is directly sensitive to the
pseudoscalar one.  At the precision of $10^{-3}$, $R_{e/\mu}^\pi$ probes
the pseudoscalar and axial vector mass scales up to 1,000\,TeV and
20\,TeV, respectively \cite{Cam05,Cam08}.  For comparison,
Cabibbo-Kobayashi-Maskawa (CKM) matrix unitarity and precise
measurements of several superallowed nuclear beta decays constrain the
non-SM vector contributions to $>20\,$TeV, and scalar to $>10\,$TeV
\cite{PDG14}.  Although scalar interactions do not directly contribute
to $R_{e/\mu}^\pi$, they can do so through loop diagrams, resulting in
sensitivity to new scalar interactions up to 60\,TeV \cite{Cam05,Cam08}.
The subject was recently reviewed at length in Ref.~\cite{Bry11}.  In
addition, $(R_{e/\mu}^{\pi})^{\rm exp}$ provides limits on masses of
certain SUSY partners \cite{Ram07}, and on neutrino sector
anomalies \cite{Loi04}.

\subsection{The PEN experiment at PSI}

\begin{figure}[!t]
    \centerline{
       \includegraphics[width=0.8\linewidth]{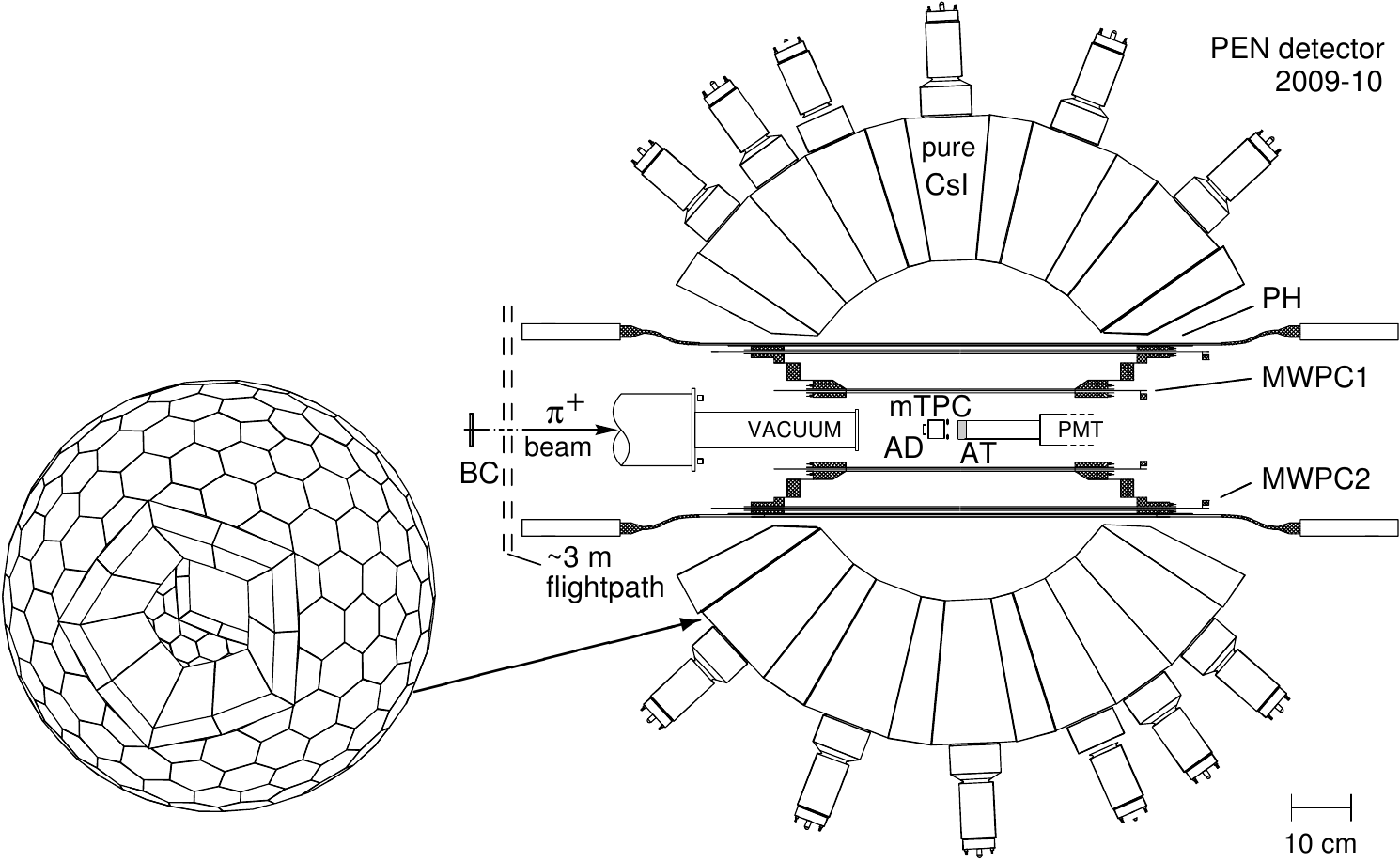}
               }
    \caption{Schematic cross section of the PEN apparatus: upstream beam
      counter (BC), 5\,mm thick active degrader (AD), mini time
      projection chamber (mTPC), active target (AT), cylindrical
      multiwire proportional chambers (MWPC's), plastic hodoscope (PH)
      detectors and photomultiplier tubes (PMT's), 240-element pure CsI
      electromagnetic shower calorimeter and its PMT's.  BC, AD, AT and
      PH detectors are made of plastic scintillator.  For details on
      detector performance see \cite{Frl04a}.}
      \label{fig:PEN_det}
\end{figure}

Between 2008 and 2010, PEN, a collaboration of 7 institutions from USA
and Europe, has carried out measurements of $\pi^+$ and $\mu^+$ decays
at rest at the Paul Scherrer Institute (PSI) with the aim to reach $
\Delta R_{e/\mu}^{\pi} / R_{e/\mu}^{\pi} \simeq 5 \times 10^{-4} $, and
is currently analyzing the data \cite{PENweb}.  The PEN experiment uses
an upgraded version of the PIBETA detector system, described in detail
in Ref.~\cite{Frl04a}, and previously used in a series of rare pion and
muon decay measurements \cite{Poc04,Frl04b,Byc09,Poc14}.  The main
component of the PEN apparatus, shown in Fig.~\ref{fig:PEN_det},
\begin{figure}[!hbt]
  \centerline{
  \parbox{0.44\linewidth}{
          \includegraphics[width=\linewidth]{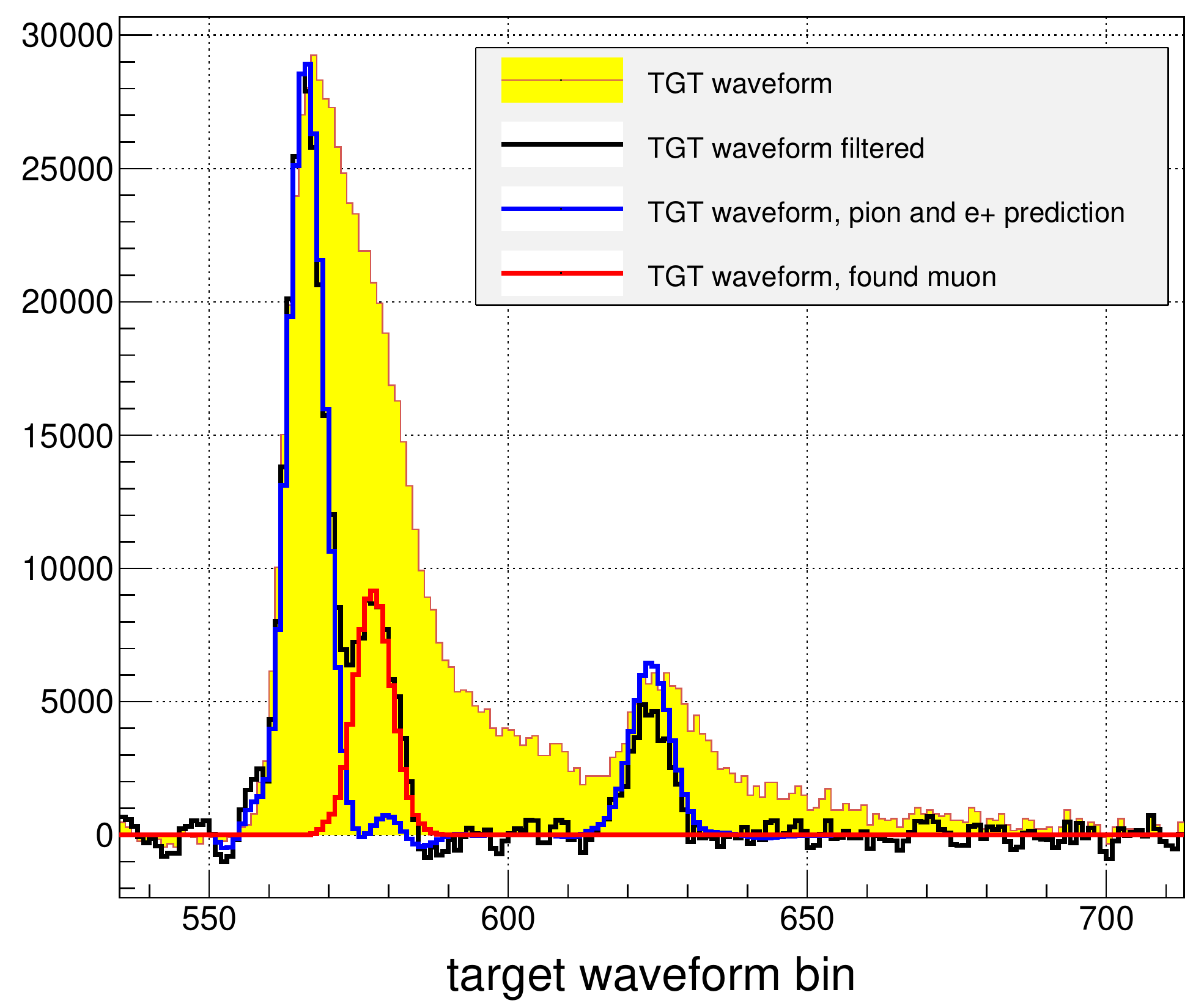}
                         }
  \parbox{0.44\linewidth}{
          \includegraphics[width=\linewidth]{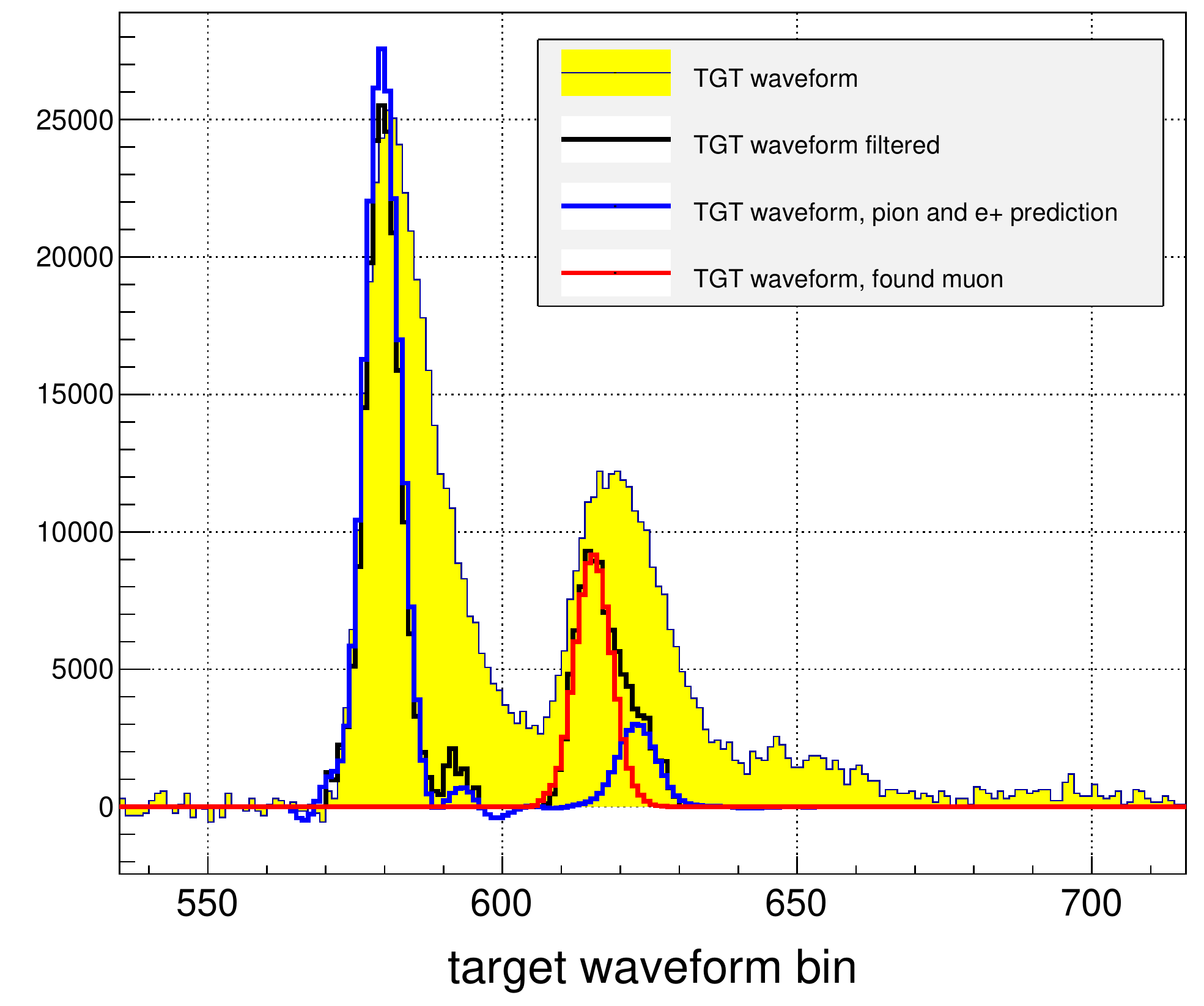}
                         } 
             }
  \caption{Full and filtered active target (TGT) waveform in the PEN
    experiment for two challenging $\pi\to\mu\to e$ sequential decay
    events with an early $\pi\to\mu$ decay (left) and early $\mu\to e$
    decay (right).  The filtering procedure consists of a simple
    algebraic manipulation of the signal.  To the naked eye both raw
    waveforms appear to have two peaks only.  The separation of events
    with/without a muon signal depends critically on the accuracy of the
    predicted pion and positron signals.
                 }  \label{fig:wf_fits}
\end{figure}
is a spherical large-acceptance ($\sim\,3\pi$\,sr) electromagnetic
shower calorimeter.  The calorimeter comprises 240 truncated pyramids of
pure CsI, 12 radiation lengths (r.l.) deep.  Beam particles entering the
apparatus with $p\simeq 75$\,MeV/$c$ are tagged in a thin upstream beam
counter (BC) and, after a $\sim 3$\,m long flight path in a 5\,mm thick
active degrader (AD) and a low-mass mini time projection chamber (mTPC),
finally to reach a 15\,mm thick active target (AT) where the beam pions
stop.  Decay particles are tracked non-magnetically in a pair of
concentric cylindrical multiwire proportional chambers (MWPC1,2) and an
array of twenty 4\,mm thick plastic hodoscope detectors (PH), all
surrounding the active target.  The BC, AD, AT and PH detectors are all
made of fast plastic scintillator material. 
Detector waveforms are digitized at
2\,GS/s for BC, AD, and AT, and at 250\,MS/s for the mTPC.
\begin{figure}[!b]
  \centerline{
   \includegraphics[width=0.75\linewidth]{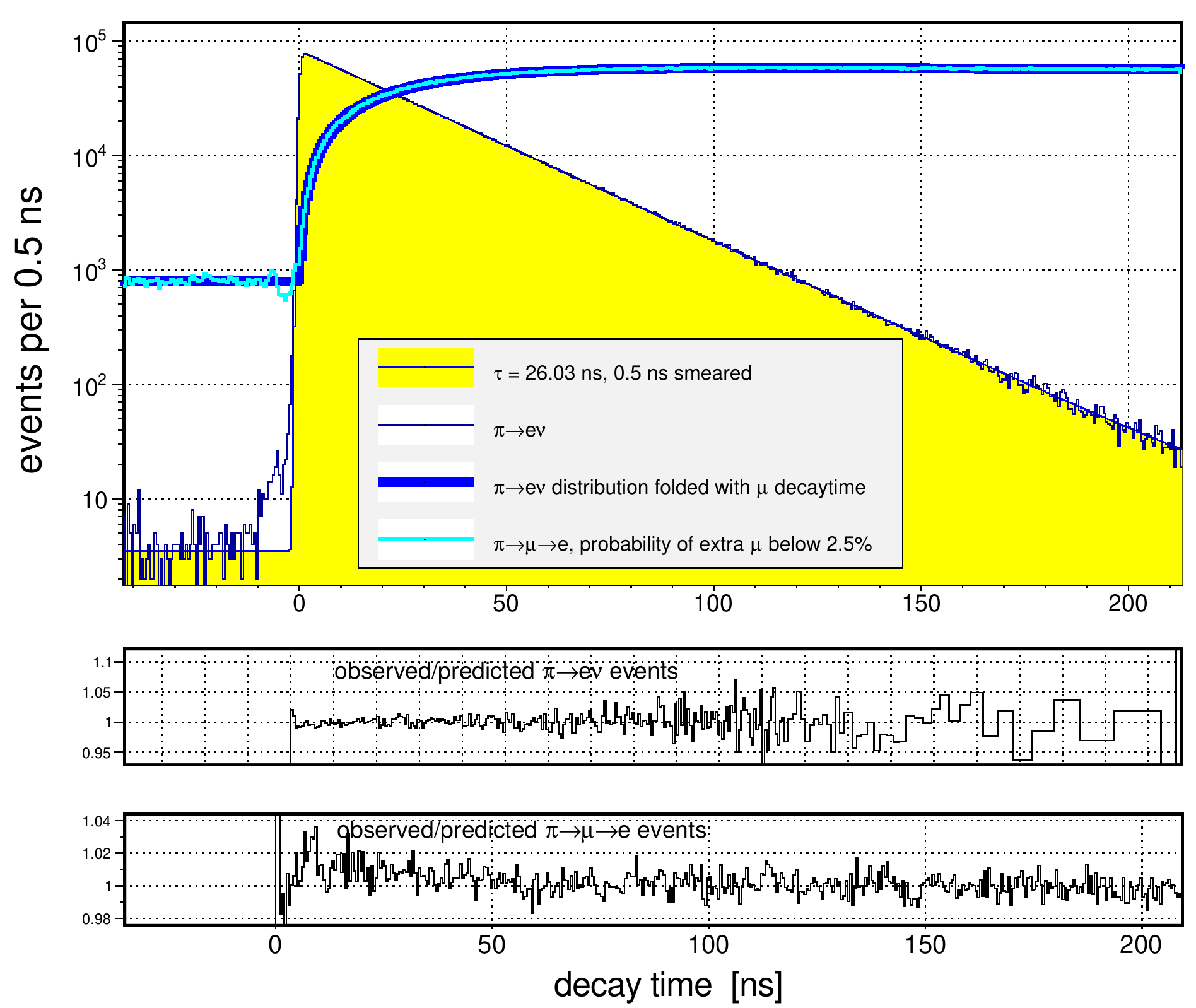}
             }
  \caption{Decay time histograms for a subset of 2010 PEN data: $\pi\to
    e\nu$ and $\pi\to\mu\to e$ events.  The two processes are
    distinguished primarily by the total $e^+$ energy and by the absence
    or presence, respectively, of an extra 4.1\,MeV (muon) in the target
    due to $\pi\to\mu$ decay.  The $\pi_{e2}$ data are shown with a pion
    lifetime $\tau_\pi=26.03$\,ns exponential decay function
    superimposed.  The $\pi\to\mu\to e$ data, prescaled by a factor of
    $\sim$\,1/64, are shown with the cut on the probability of $< 2.5$\%
    for a second, pile-up muon to be present in the target at
    $t=0=t_{\pi\text{stop}}$.  The turquoise histogram gives the
    $\pi\to\mu\to e$ yield constructed entirely from the measured
    $\pi\to e\nu$ data folded with the $\mu$ decay rate, and corrected
    for random muons; it perfectly matches the bold dark blue histogram.
    The two lower plots show the observed to predicted ratios for
    \peii\ and $\pi\to\mu\to e$ events, respectively; the observed
    scatter is statistical in nature. }
    \label{fig:pen-decaytime}
\end{figure}

A key source of systematic uncertainty in \peii\ measurements at rest
has been the hard to measure low energy tail of the detector response
function, caused by electromagnetic shower leakage from the calorimeter
mostly in the form of photons, masked by the overwhelming $\pi\to\mu\to
e$ background events.  Other physical processes, if not properly
identified and suppressed, also contribute events, mainly to the low
energy part of the spectrum.  They include: ordinary pion decay into a
muon in flight, before the pion is stopped, with the resulting muon
decaying within the time gate accepted in the measurement, and radiative
decay events.  The latter process is well measured, analyzed, and
properly accounted for in the PEN apparatus.  Shower leakage and pion
decays in flight can only be well characterized if the $\pi\to\mu\to e$
chain can be well separated from the direct $\pi \to e$ decay in the
target.  Therefore much effort has been devoted to digitization,
filtering and analysis of the target waveforms \cite{Pal12}, as
illustrated in Fig.~\ref{fig:wf_fits}.  The decay time histograms of the
$\pi\to e\nu$ decay and $\pi\to \mu\to e$ sequence, shown in
Fig.~\ref{fig:pen-decaytime} for a subset of data recorded in 2010,
illustrate best the quality of the PEN data.  The $\pi\to e\nu$ data
follow the exponential decay law over more than three orders of
magnitude, and perfectly predict the measured $\pi\to\mu\to e$
sequential decay data once the latter are corrected for random (pile-up)
events.  Both event ensembles were obtained with minimal requirements
(cuts) on detector observables, none of which bias the selection in ways
that would affect the branching ratio.  The probability of random
$\mu\to e$ events originating in the target can be controlled in the
data sample by making use of multihit time to digital converter (TDC)
data that record early pion stop signals.  With this information one can
strongly suppress events in which an ``old'' muon was present in the
target by the time of the pion stop that triggered the readout.

During the 2008-10 production runs the PEN experiment accumulated some
$2.3 \times 10^7$ $\pi\to e\nu$, and more than $2.7 \times 10^8$
$\pi\to\mu\to e$ events, as well as significant numbers of pion and muon
radiative decays.  A comprehensive blinded maximum likelihood analysis
is under way to extract a new experimental value of $R_{e/\mu}^{\pi}$.
As of this writing, there appear no obstacles that would prevent the PEN
collaboration to reach a precision of $\Delta R/R < 10^{-3}$.  The PIENU
experiment at TRIUMF, discussed below, has a similar precision goal.

\subsection{The PIENU experiment at TRIUMF}

\begin{figure}[!b] 
  \includegraphics[width=0.5\linewidth]{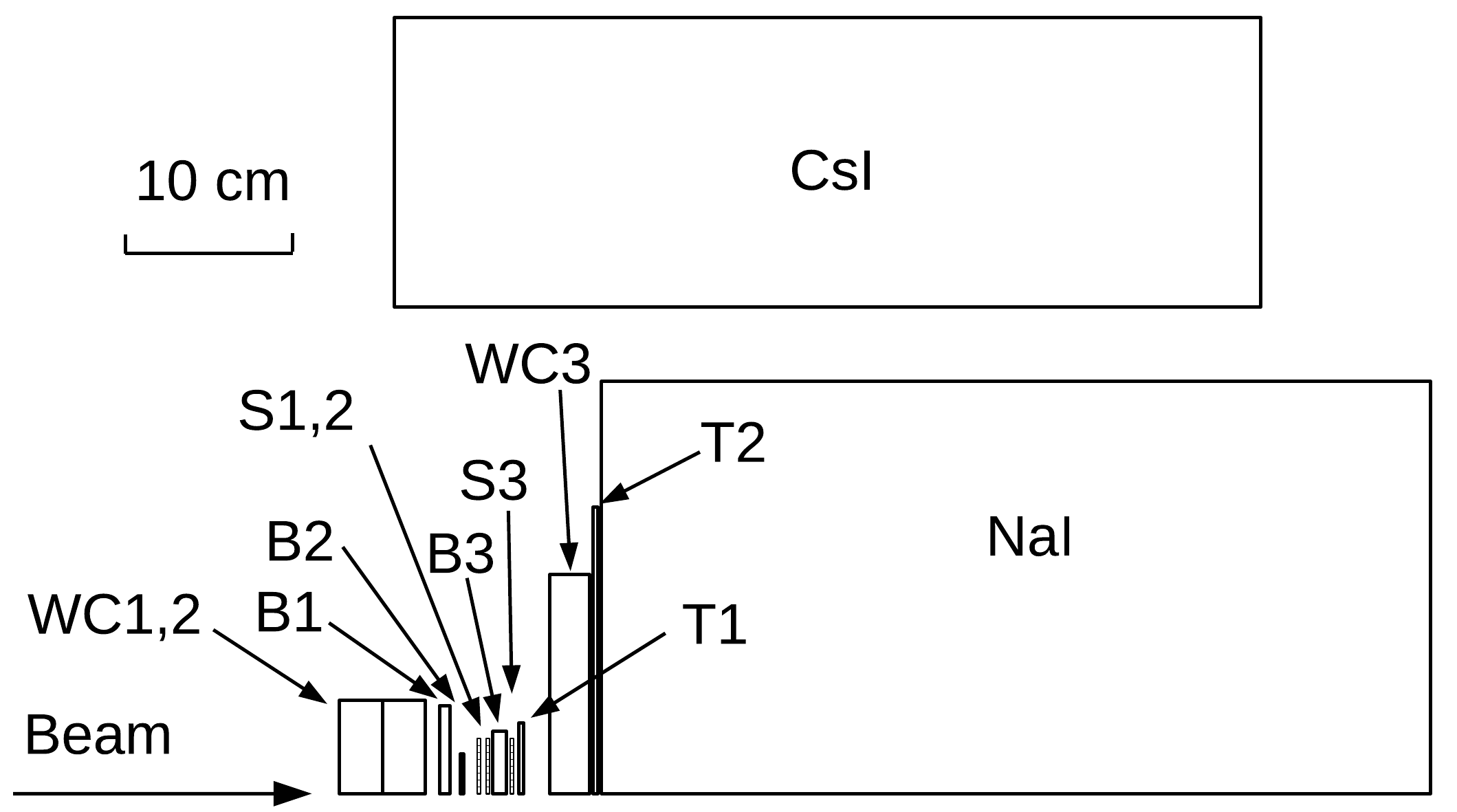}
  \caption{Top half cross-section of the PIENU detector. The cylindrical
    NaI(Tl) crystal is surrounded by a cylindrical array of CsI
    crystals.  For further details see Ref.~\cite{Agu15} and references
    therein.} 
  \label{fig:PIENU_det}
\end{figure}

The PIENU experiment at TRIUMF builds on the earlier measurements at the
same laboratory \cite{Bri92}, aiming at a significant improvement in
precision through refinements of the technique used.  Major improvements
in precision in PIENU over the earlier TRIUMF TINA measurement derive
from improved geometry and beamline, a superior calorimeter, as well as
high-speed digitizing of all detector signals.  The apparatus is
described in detail in Ref.~\cite{Agu15} and shown in
Fig.~\ref{fig:PIENU_det}.  
A 75\,MeV/$c$ $\pi^+$ beam from the improved
TRIUMF M13 beam line \cite{Agu09} is tracked in wire chambers,
identified by plastic scintillators, and stopped in a 0.8\,cm thick
scintillator target.  Fine tracking near the target is provided by two
sets of single-sided silicon strip detectors located immediately
upstream and downstream of the target assembly.  The positrons from
$\pi\to e\nu$ and $\pi\to\mu\to e$ decays are detected in the positron
telescope, which consists of a silicon strip counter, two thin plastic
counters, and an acceptance-defining wire chamber that covers the front
of the crystal calorimeter.  The calorimeter is 19 r.l. deep and
consists of a large single crystal of NaI(Tl) surrounded laterally by an
array of 97 pure CsI crystals.  The solid-angle acceptance of the
telescope counters is 20\% of 4$\pi$\,sr.  Thus, PIENU achieves better
energy resolution than PEN, at the expense of lower solid angle
coverage.  The PIENU experiment completed data acquisition in 2012, and
is in the analysis stage as of this writing.
The PIENU collaboration has recently published an analysis of 1/5 of
their data \cite{Agu15} with the result
\begin{equation}
   \left( R^{\pi}_{e/\mu}\right)^{\text{PIENU}} = 
         1.2344\,(23)_{\text{stat}}\,(19)_{\text{syst}} \times 10^{-4}\,,
\end{equation}
which is consistent with the previous world average \cite{PDG14} as well
as with the Standard Model prediction of Eq.~\ref{eq:pi_e2_full_SM}, and
has the effect of reducing the previous world average uncertainty by almost
a factor of two, as shown in Eq.~\ref{eq:pi_e2_2015_avg}.



\section{\boldmath Pion radiative electronic $\pi^+\to e^+\nu\gamma$ decay
                   (\peiig)}

The decay $\pi^+ \to e^+\nu_e\gamma$ proceeds via a combination of QED
(inner bremsstrahlung, $IB$) and direct, structure-dependent ($SD$)
amplitudes \cite{Don92,Bry82}.  The strong helicity suppression of the
primary non-radiative process, $\pi\to e\nu$, discussed above, also
suppresses the $IB$ terms, making the structure-dependent amplitudes
measurable in certain regions of phase space \cite{Don92,Ber13}.  To
describe the $SD$ amplitude, standard \VmA\ electroweak theory requires
only two pion form factors, $F_A$, axial vector, and $F_V$, vector (or
polar-vector).
\begin{figure}[!b]
  \centerline{\includegraphics[width=0.5\linewidth]{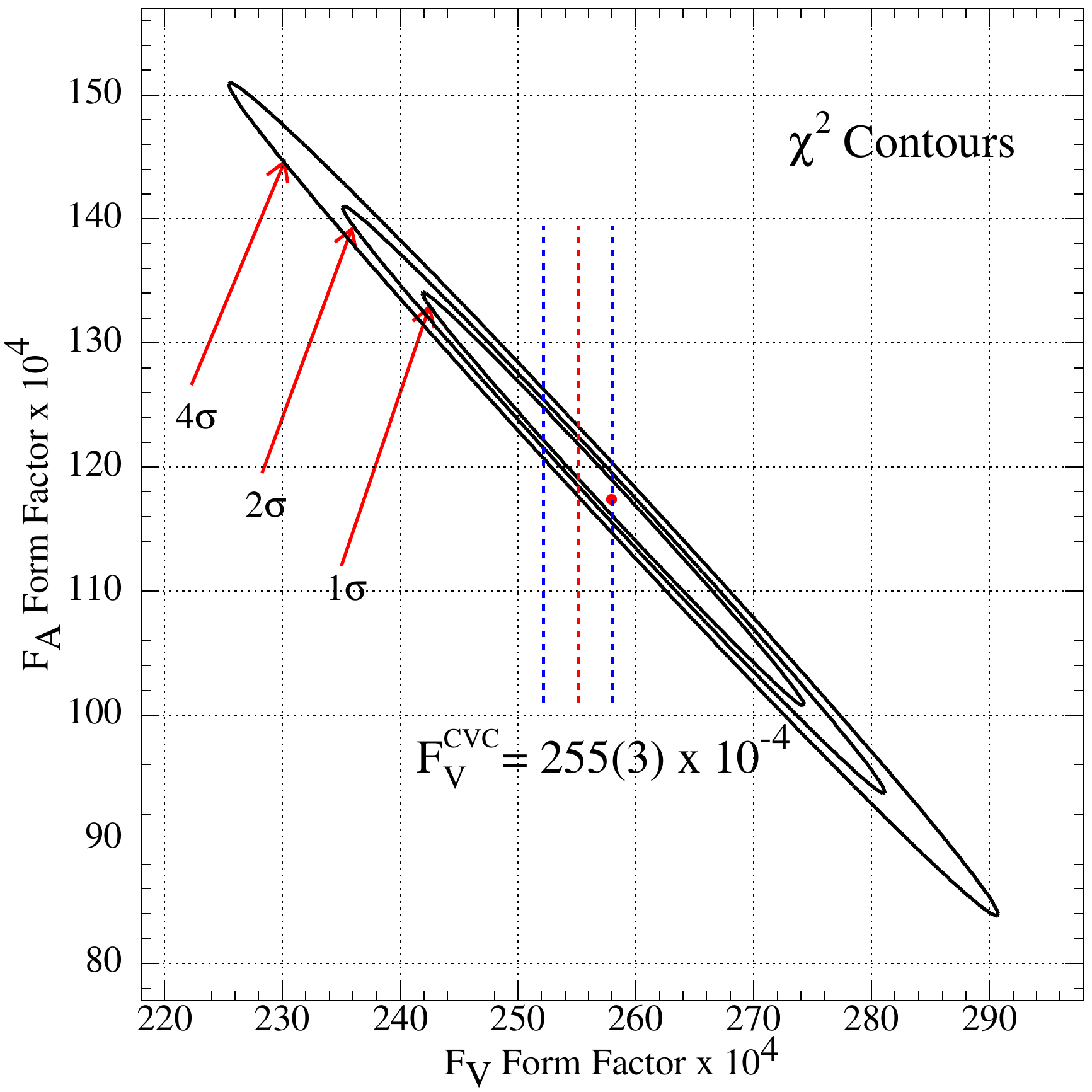} }
  \caption{PIBETA data: contour plot of loci of constant $\chi^2$ for
    the minimum value $\chi^2_0$ (red dot) plus 1, 2, and 4 units,
    respectively, in the $F_A$-$F_V$ parameter plane.  The range of the
    CVC prediction $F_V = 0.0255 \pm 0.0003$ is indicated by the dashed
    vertical lines.  See Refs.~\cite{Byc09,Poc14} for more details.}
  \label{fig:F_A-F_V-2013}
\end{figure}
The amplitudes $F_A$ and $F_V$ in principle depend on the 4-momentum
transfer $q^2$ to the $e$-$\nu$ pair (or to the $W$ boson); in
\peiig\ decay $q^2\approx 0$ is a good approximation (``soft pion
limit'').  For a long time radiative pion decay measurements had access
only to one structure dependent amplitude, the $SD^+ \propto
(F_V+F_A)^2$, with weak or no sensitivity to $SD^- \propto (F_V-F_A)^2$.
Therefore most evaluations took the value of $F_V$ from the conserved
vector current (CVC) hypothesis prediction based on the $\pi^0 \to
\gamma\gamma$ decay width \cite{Don92}.  Recent PIBETA collaboration
results \cite{Byc09} led to an order of magnitude improvement in the
precision of the \peiig\ branching ratio determination, as well as of
$F_A$ and $F_V$, and a first evaluation of the $q^2$ dependence of
$F_V$.  However, the measurement was most sensitive to the low $p_\nu$
segment of phase space which is strongly dominated by the $SD^+$
amplitude, resulting in a very narrow constraint on $F_V+F_A$, as shown
in Fig.~\ref{fig:F_A-F_V-2013}.

The PEN data analysis is expected to add significantly to the more than
60\,k PIBETA \peiig\ event set, but with increased sensitivity to the
$SD^-$ amplitude due to better controlled backgrounds.  Thus, the
extremely skewed ellipse of Fig.~\ref{fig:F_A-F_V-2013} would give way
to a more balanced set of limits, and, thus, an improved independent
limit on $F_V$, as well as a further tightening of the limit on $F_T$,
the long hinted-at tensor contribution \cite{Byc09}.  It is worth noting
that the ratio of $F_A/F_V$ enters directly into the chiral perturbation
theory lagrangian at the leading order through the $l_9+l_{10}$ term
\cite{Don92}, and is among the basic low energy chiral constants.

\section{\boldmath Pion semileptonic (beta) $\pi^+\to \pi^0e^+\nu$ decay
                   (\peiii)}

Unlike \peii, the extremely rare, ${\cal O}(10^{-8})$, pion beta decay
is not suppressed; its low rate derives from the restricted phase space
of final states, entirely due to the small difference between the
$\pi^\pm$ and $\pi^0$ masses.  As a pure vector $0^-\to 0^-$
transition, it is fully analogous to the superallowed Fermi (SAF)
nuclear beta decays; indeed it is the simplest realization of the
latter, fully free of complications arising from nuclear structure
corrections.  SAF decays have historically led to the formulation of the
CVC hypothesis, and have played a critical role in testing the unitarity
of the Cabibbo-Kobayashi-Maskawa quark mixing matrix through evaluations
of the $V_{ud}$ element \cite{PDG14}.  

The $\sim$\,0.5\,\% PIBETA \peiii\ measurement \cite{Poc04} is the most
precise one to date.  Because it used \peii\ decay events for
normalization, this result will receive a slight improvement in
precision once the PEN and PIENU results become available.  Although not
competitive with the SAF based $V_{ud}$, there are no plans to improve
the PIBETA result precision until the current crop of experiments
studying the more easily accessible neutron beta decay are completed
(for a more detailed discussion of that topic see Ref.~\cite{Bae14}).
In the meantime, however, one can use the PIBETA $\pi_{e3}$ branching
ratio to evaluate $R_{e/\mu}^\pi$ by fixing $V_{ud}$ to its very precise
PDG 2014 recommended value of $0.97425\,(22)$ \cite{PDG14} and adjusting
$R_{e/\mu}^\pi$ until the extracted value of $V_{ud}^{\pi\beta}$ agrees.
This exercise yields:
\begin{equation}
   (R_{e/\mu}^\pi)^{\rm PIBETA} = 1.2366\,(64) \times 10^{-4}\,,
\end{equation}
in good agreement with direct measurements reviewed in the above section
on \peii\ decay.

\section{\boldmath Muon radiative $\mu^+\to e^+\nu\bar{\nu}\gamma$
  decay} 

\begin{figure}[!htb]
 \centerline{
  \parbox{0.44\linewidth}{
    \includegraphics[width=\linewidth]{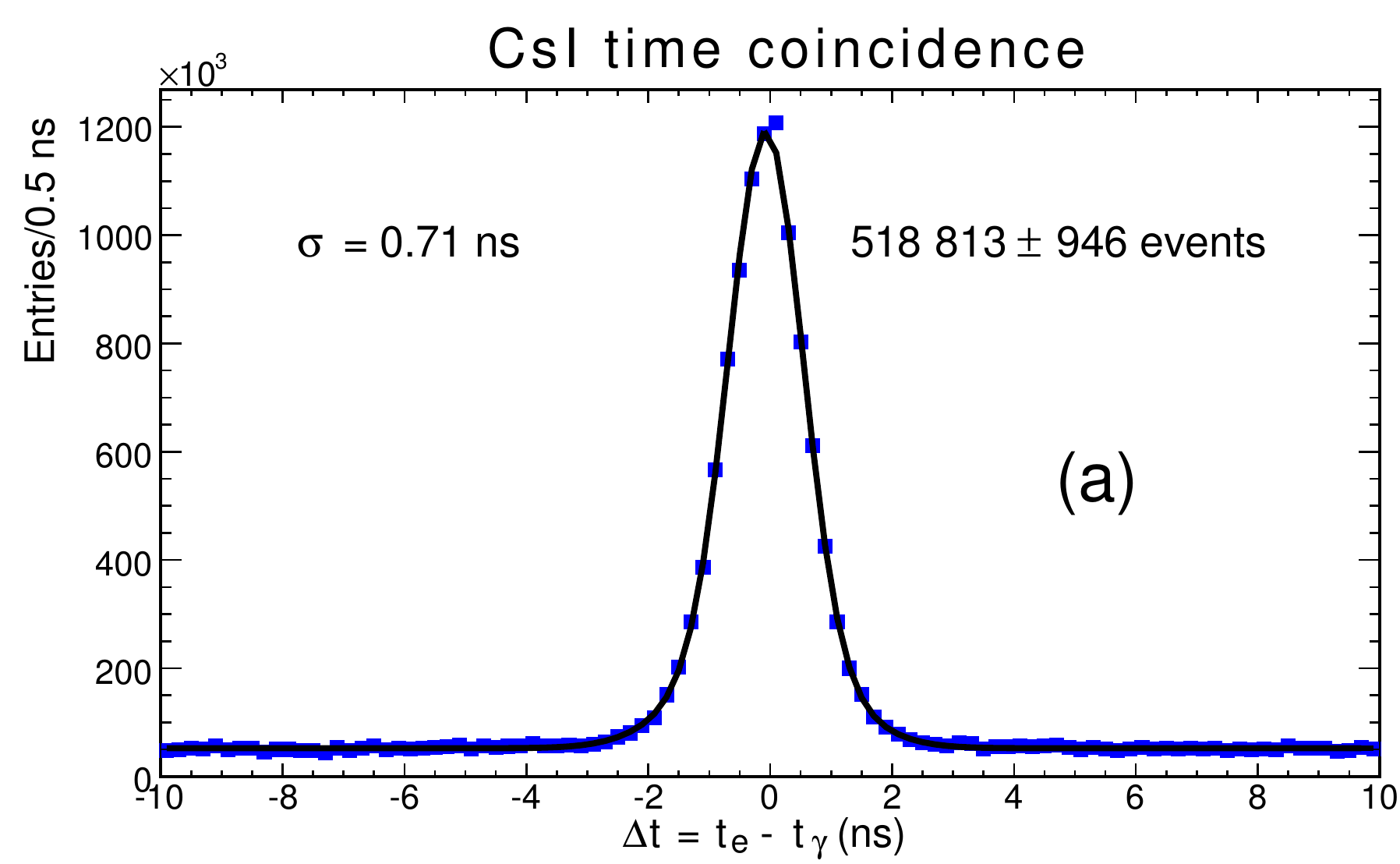}
                         }
  \parbox{0.44\linewidth}{
  \includegraphics[width=\linewidth]{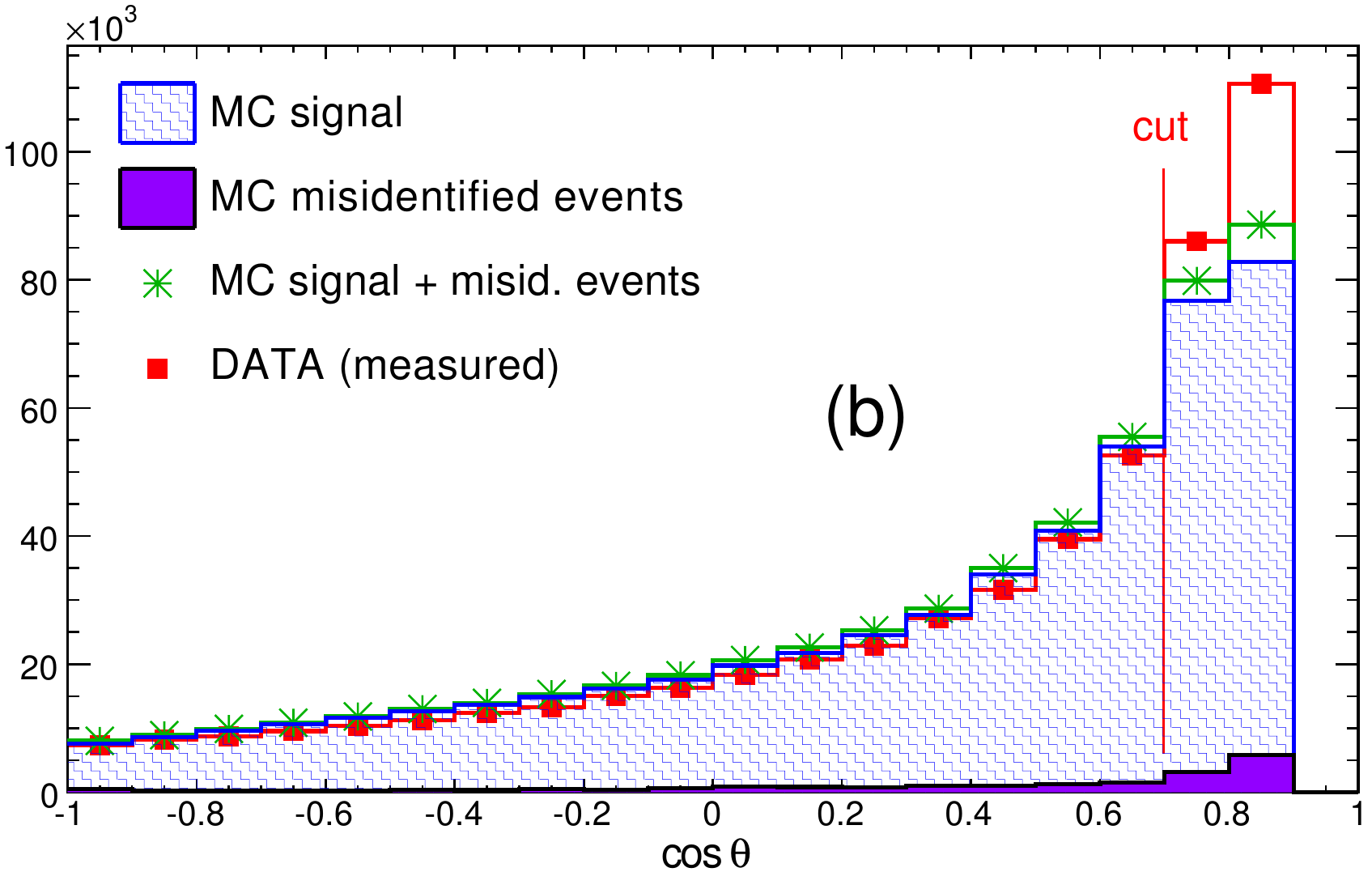} 
                         }
            }
 \centerline{
  \parbox{0.44\linewidth}{
    \includegraphics[width=\linewidth]{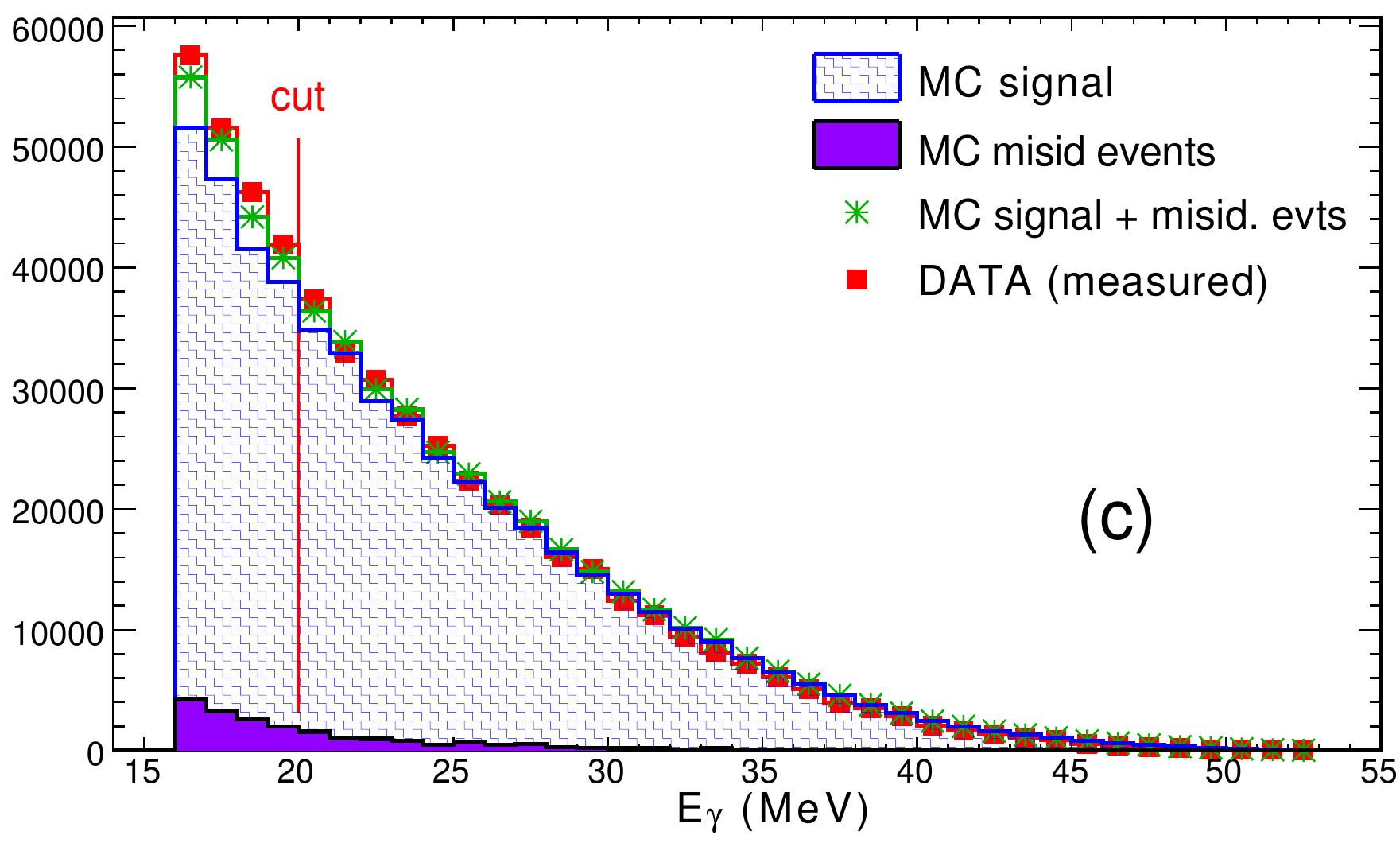}
                         }
  \parbox{0.44\linewidth}{
  \includegraphics[width=\linewidth]{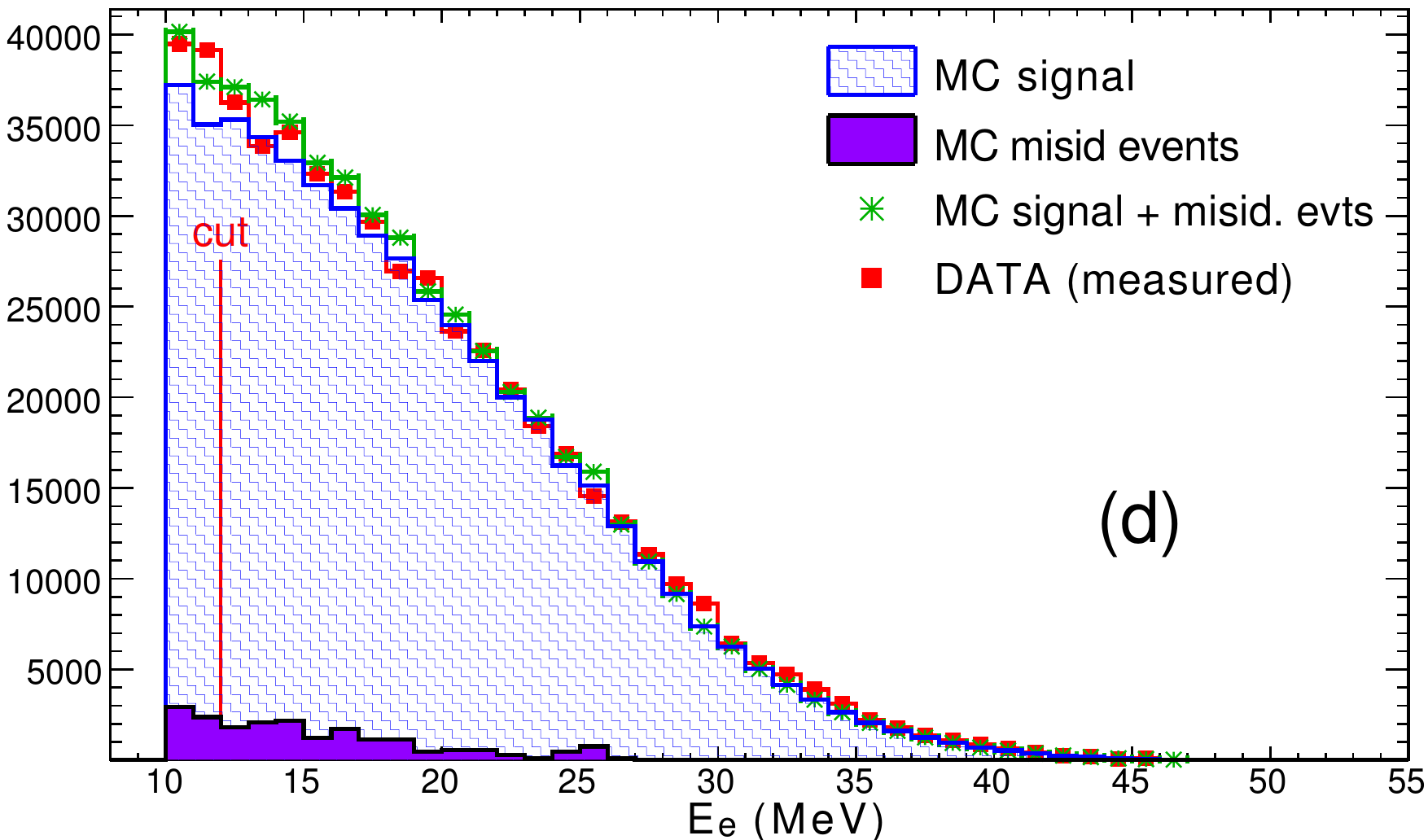}
                         }
            }
  \caption{PIBETA data: measured and simulated $\mu^+\to
    e^+\nu_e\bar{\nu}_\mu\gamma$ distributions of (a) $\Delta
    t_{e\gamma}$, (b) $\cos\theta_{e\gamma}$, (c) $E_\gamma$, and (d)
    $E_{e^+}$.  Also shown are the misidentified Monte Carlo events
    (split-off secondary neutral showers), as well as bounds of cuts
    applied in the branching ratio analysis.
    \label{fig:rmd}}
\end{figure}

A 2004 PIBETA set of $\sim$0.5\,M radiative muon events was recently
analyzed; the relevant measured and Monte Carlo simulated spectra,
including backgrounds, shown in Fig.~\ref{fig:rmd}, are in excellent
agreement within the design acceptance of the spectrometer.  The
analysis yields a preliminary branching ratio for $E_\gamma >
10$\,MeV, and $\theta_{e\gamma} > 30^\circ$:
\begin{equation}
  B^{\text{exp}}(\mu^+ \to e^+\nu_{\text{e}}\bar{\nu}_\mu\gamma)  =
      4.365\,(9)_{\text{stat}}\,(42)_{\text{syst}}  \times 10^{-3}\,,
   \label{eq:B-RMD}
\end{equation}
which represents a 29-fold improvement in precision over the previous
result \cite{PDG14}, and is in excellent agreement with the SM value:
$B^{\text{SM}} = 4.342\,(5)_{\text{stat-MC}} \times 10^{-3}$.
Minimum-$\chi^2$ analysis of the most sensitive data subset (with
roughly balanced systematic and statistical uncertainties) yields a
preliminary value for the $\bar{\eta}$ parameter
($\bar{\eta}^{\text{SM}}\equiv 0$):
\begin{equation}
  \bar{\eta} =  0.006\,(17)_{\text{stat}}\,(18)_{\text{syst}},\qquad
      \text{or} \qquad \bar{\eta}  < 0.028 \quad (68\% \text{CL})\,,
\end{equation}
a 4-fold improvement over previous limits \cite{Eic84}.  Details of this
analysis, including a discussion of the uncertainties, are given in
Refs.~\cite{Mun12} and \cite{Van05}.

\section{Conclusions}\vspace*{-6pt}

Vigorous efforts are presently under way to measure precisely the
branching ratios for allowed rare decays of the charged pion as well as
of the muon.  The experimental precision still lags by about an order of
magnitude behind SM calculations.  As that gap is narrowed, this field
of research, complementary to collider searches, will realize its full
potential for discovery or further improvement of the limits on various
possible extensions of the Standard Model beyond the well established
\VmA\ form.  Specifically, a significant improvement of the precision of
the \peii\ branching ratio is expected from the full PEN and PIENU
analyses which are forthcoming in the near future, with attendant limits
on lepton universality and non-(\VmA) interaction terms.  

\begin{acknowledgments}

This work has been supported by grants from the US National Science
Foundation, the Paul Scherrer Institute, and the Russian Foundation for
Basic Research.

\end{acknowledgments}


\end{document}